# On Dark Energy and Dark Matter (Part II)

Shlomo Barak and Elia M. Leibowitz
School of Physics & Astronomy, Tel Aviv University
22 October 2008

## Abstract

Phenomena currently attributed to Dark Matter (DM) and Dark Energy (DE) are merely a result of the interplay between gravitational energy density, caused by the contraction of space by matter, and space dilation, caused by the energy density of the Cosmological Microwave Background (CMB).

This interplay causes inhomogeneous and anisotropic expansion, in and around galaxies, whereas the expansion of the universe, when viewed globally, is homogeneous and isotropic.

These contentions lead to a theoretical derivation of the gravitational central acceleration in and around galaxies, and the determination of $g_0$, the central acceleration where flattening of Rotation Curves (RC) replaces Keplerian behavior.

Our results, which fit the observed flattening of RCs, resemble the phenomenological Tully-Fisher and Millgrom MOND relations. However, our central acceleration, $g_0$, depends on the CMB energy density at the time of formation of a galaxy and, as opposed to MOND, is **not** a universal constant.

---

This paper is divided into three parts. Part I (to be published) discusses DE, Cosmology and explains our contention that the CMB energy density dilates space, which implies that electromagnetism is a non-linear phenomenon, similar to the non-linear phenomenon of gravitation. Part II (this paper) addresses the DM and astrophysical issues. Part III will introduce a model that describes the formation and evolution of galaxies and their resulting Rotation Curves (RC).

The ideas and notions raised in this trilogy are more fully discussed in a forthcoming book by one of us, (S. Barak, 2009) that reconsiders the foundations of Physics.

## Dark Matter

To address the issue of DM, we examine basic concepts in cosmology.

1. **Space is three-dimensional, foamy, elastic and vibrating.**
   The assumption that space is 3D means that the universe is **not** a curved 3D manifold in a hyperspace with an additional spatial dimension. For a flat universe the issue of an additional spatial dimension is not relevant.
   The idea, shared by many, that space is foamy, and hence cellular, rests on the meaning of expansion, and the requirement that its vibrations have a finite energy density.
   In this discussion, curved space means space that is elastically deformed. This deformation is the change in size of the space cells. Positive or negative curvature, around a point in space, means that the space cells grow or shrink, respectively, from this point outwards. For a positively curved manifold, the ratio of the circumference of a circle to the radius is less than $2\pi$, as measured by a yardstick of a fixed length. For an elastically positively curved 3D-space around a point the above ratio is also less than $2\pi$, as measured by a flexible yardstick such as a space cell. Note that for an **elastic space** there is no meaning to global curvature, **curvature is a local attribute**. The smallest linear dimension of a space cell, whether Planck's length or not, which determines the Zero Point Fluctuations (ZPF) energy density, is not relevant to our discussion here. We suggest that space vibrations are the electromagnetic waves and that these waves dilate space due to its anharmonicity in a similar manner to thermal (vibrational) energy in solids. Hence, globally, the CMB is the main contributor to space dilation over the average space density set by the ZPF, (A. D. Sakharov, 1968 and C. W. Misner et al, 1970). As such, we can consider the effect of the CMB to be that of an anti-gravitational cosmological "constant".

$\epsilon = mc^2$ does not imply that all forms of energy curve space in the same way. We contend that energy in the form of matter curves space positively by contracting it. However, energy in the form of electromagnetic waves dilates space and thus can contribute, locally, to positive curving around a mass, or globally to flattening.

The above has **implications** for both the mass equivalence principle and the way that the energy densities of gravitation and electromagnetic waves are incorporated in the General Relativity (GR) field equation. This is discussed in Part I and in a book by S. Barak (2009).

2. **Space contraction or expansion is the change of cell sizes.**
   GR shows that a mass **contracts** space around it. Cells close to the mass are smaller than those at a distance and hence the elastic **positive** curving of space. Length, close to a mass, is smaller, and the running of time is slower, than at a distance. Space expansion is the enlargement of its cells, and the CMB dilating vibrational energy contributes to this expansion.
   Note that Riemannian geometry allows two types of 3D curved space, a 3D curved manifold within a 4D hyperspace with an extra spatial dimension, or an elastic 3D space, which we suggest is the reality for both astrophysical and cosmological phenomena.

3. **In and around galaxies, space is deformed by an inhomogeneous expansion.**
   This deformation depends on gravitational contraction, due to mass, expressed by gravitational energy density, $\epsilon_g$, and the opposing dilation, due to the dilating vibrational CMB density, $\epsilon_{CMB}$. We contend that wherever $\epsilon_g > \epsilon_{CMB}$ space expansion is inhibited.
   To make the theoretical basis for our discussion as clear as possible, we use a **simplified** model of a galaxy.
   We consider a "point" mass galaxy whose formation time (the mass accretion phase) is much shorter than its present age. In other words, we assume that the galaxy was formed "instantly" at time $t_0$, when the scale factor was $a_0$, possessing its final mass value. The redshifted galactic light recorded now left the galaxy at cosmic time $t_z$, when the scale factor was $a_z$. Paper III presents a more realistic case in which evolution is taken into account. It calculates RCs, based on the theory presented here, and compares them with the observed RCs of real galaxies.

For simplicity, we divide the space around a galaxy into three regions according to the relative strengths of $\epsilon_g$ and $\epsilon_{CMB}$:

a. **Close to the center of a galaxy, where $\epsilon_g > \epsilon_{CMB}$.**
   In this region, the local contraction of space by the mass of the galaxy is stronger than the opposing dilation caused by the CMB. Space expansion is inhibited in this region, and hence Keplerian behavior is observed.

b. **Further from the center of a galaxy, from $R_0$ to R as defined below, where the relationship $\epsilon_g \approx \epsilon_{CMB}$ is maintained for decreasing values of $\epsilon_g$ and $\epsilon_{CMB}$.**
   In this region, equilibrium was first attained at a distance $R_0$, at the time, $t_0$, of formation of the galaxy. The expansion of the surrounding space beyond $R_0$, due to the expansion of the universe, lowered the $\epsilon_{CMB}$, and hence equilibrium was reached for $t > t_0$, at a greater distance $r(t) > R_0$. This is an ongoing process in which the region surrounding $R_0$ grows with time, with an ever-increasing value of the scale factor.
   Light that reaches us now, left the galaxy at time $t_z$. Equilibrium, $\epsilon_g \approx \epsilon_{CMB}$, at this time, occurred at a distance R from the center of the galaxy. Space density in the region between $R_0$ and R is **frozen**, since $\epsilon_g > \epsilon_{CMB}$. Space density at $R_0$ is larger than at R. In this region, flattening of RCs replaces Keplerian behavior.

c. **Far from the galaxy, where $\epsilon_g < \epsilon_{CMB}$.**
   In this region, Keplerian behavior is again observed.

As we show, this inhomogeneous expansion is mistakenly interpreted as DM. To express this inhomogenity we introduce a scale factor a(r,t) that depends not only on time, t, but also on the distance, r, from the center of a galaxy.

4. **We modify the Newtonian gravitational field equation by taking into account space deformation due to its expansion.**
   Newtonian gravitation is an approximation since, unlike GR, it does not take into account, in the calculation of the flux density, the curving of space around a mass. The terms "curved" and "deformed" are used interchangeably in this paper.
   Let d be the distance between two points at a distance r from the center of a galaxy, as measured by the fixed yardstick of an observer located at the distance $R_0$, where the scale factor is $a(R_0,t)$. For an observer with a fixed yardstick at a distance $r > R_0$ from the center, where the local scale factor is a(r,t), the measured distance is d**,** where**:**

$$d' = d \cdot \frac{a(R_0,t)}{a(r,t)}$$

In space that is contracted gravitationally: $d' < d$ since $\frac{a(R_0,t)}{a(r,t)} < 1$

This also holds for space that is both contracted gravitationally and is expanded inhomogeneously with r.

The area of a virtual spherical shell of radius r in deformed space, with the scale factor, a(r,t), is $A'$. The area of a spherical shell of the same radius in un-deformed space, in which the value of the scale factor, a($R_0$,t), is uniform, is A, where A′ is related to A as follows:

$$A' = A \cdot \left[\frac{a(R_0,t)}{a(r,t)}\right]^2$$

Therefore, in an expanding space, the Gauss theorem implies that the field strength, $E_g(r,t)$, which is the flux density perpendicular to the shell, is larger than the field strength for un-deformed (flat) space:

$$(1) \quad E_g(r,t) = \frac{GM}{r^2}\left[\frac{a(r,t)}{a(R_0,t)}\right]^2$$

This is **our modified universal Newtonian Gravitational field equation**.
By introducing the scale factor, a(r,t), equation (1) takes into account the curving of space by mass - by its direct contraction of space as expressed by GR as well as by its effect in modifying space expansion. The contribution of space expansion to the variation of the scale factor as a function of r is orders of magnitude greater than the intrinsic contribution by the mass, as expressed by GR. Therefore equation (1) is more applicable, in the case of inhomogeneous space expansion, than GR.
$R_0$ is the distance from the center of a galaxy at which $\epsilon_g = \epsilon_{CMB}$ at the time $t_0$ of its formation. Let $g_0$ be the central acceleration at this point. As we show, RC flattening starts at $R_0$.

Note that the broken symmetry of translation in a deformed space, and the broken symmetry of time in an expanding universe, lead to non-conservation of linear momentum and non-conservation of energy, respectively.
This implies that radiation density, like the CMB, which is homogeneous throughout space, including the interiors of "DM halos" (B. R. Granitt et al, 2008) is reduced with expansion. Far from masses, where $a(t_2) > a(t_1)$, for all $t_2 > t_1$, $\epsilon_{CMB}(t_2)$ is related to $\epsilon_{CMB}(t_1)$, as follows:

$$(2) \quad \epsilon_{CMB}(t_2) = \epsilon_{CMB}(t_1)\left(\frac{a(t_1)}{a(t_2)}\right)^4$$

5. **From the general expressions for $E_g(r,t)$ and $\epsilon_{CMB}(r,t)$, we derive the gravitational central acceleration inside galaxies for the region between $R_0$ and R.**
   **We also determine $g_0$, the central acceleration where flattening of RCs replaces Keplerian behavior.**
   Our equations resemble the phenomenological Millgrom MOND and Tully-Fisher relations.
   Consider a point in the second region, $R_0 < r < R$. From equation (1) for $E_g(r,t)$ we derive the gravitational energy density of contraction, $\epsilon_g(r,t)$ in this region:

$$(3) \quad \epsilon_g(r,t) = \frac{1}{8\pi G}E_g^2(r,t) = \frac{1}{8\pi G}\left[\frac{GM}{r^2} \cdot \left[\frac{a(r,t)}{a(R_0,t_0)}\right]^2\right]^2$$

a(r,t) is the scale factor at the distance r for all times later than, t. a($R_0$,t) is the scale factor at the distance $R_0$, which was fixed at time $t_0$, and remains the same for all times later than $t_0$. Equation (2), can thus be written for t and $t_0$ as:

(4) $\in_{CMB}(r,t) = \in_{CMB}(R_0,t_0) \cdot \left[\dfrac{a(r,t)}{a(R_0,t_0)}\right]^{-4}$

Equating equation (4) to (3) gives:

(5) $\dfrac{a(r,t)}{a(R_0,t)} = \left[\dfrac{8\pi G \in_{CMB}(R_0,t_0)}{G^2 M^2}\right]^{\frac{1}{8}} \cdot r^{\frac{1}{2}}$

We designate $E_g$ by g and the nominator in (5) by:

(6) $g_0^2 = 8\pi G \in_{CMB}(R_0,t_0)$

This designation is explained at the end of this section and in the following section. We rewrite equation (5) as:

(7) $\dfrac{a(r,t)}{a(R_0,t_0)} = \left[\dfrac{g_0^2}{G^2 M^2}\right]^{\frac{1}{8}} \cdot r^{\frac{1}{2}}$

Substituting (7) into (1) gives:

(8) $g = \dfrac{GM}{r^2} \cdot \left[\dfrac{g_0^2}{GM}\right]^{\frac{1}{4}} \cdot r = \left[(GM)^2 \cdot \dfrac{g_0}{GM}\right]^{\frac{1}{2}} \cdot r^{-1} = [(GM) \cdot g_0]^{\frac{1}{2}} \cdot r^{-1}$

Thus the **gravitational central acceleration in the region R to $R_0$** is:

(9) $g = \dfrac{\sqrt{g_0 GM}}{r}$

which, by squaring, gives the **Millgrom** MOND relation:

(10) $\dfrac{g^2}{g_0} = \dfrac{GM}{r^2}$

Since $g = \dfrac{v^2}{r}$ we get: $\dfrac{v^4}{r^2} = g_0 \dfrac{GM}{r^2}$ or:

(11) $v^4 = (g_0 G) M$ which is the **Tully-Fisher** relation.

Observations show that for many galaxies $g_0$ has the same value, $g_0 \sim 1.2 \cdot 10^{-8}$ cm sec$^{-2}$ (M. Millgrom, 2008). We contend that this is the case for galaxies formed ~ 12 BY ago, corresponding to z ~ 3, and hence to a ~ 0.25, which is the epoch of galaxy formation (C. M. Baugh et al, 1998). From equation (6), which defines $g_0$:

$\dfrac{1}{8\pi G} g_0^2 = \in_{CMB}(R_0,t_0)$

$g_0$ is thus the field (central acceleration) at $R_0$, at the time, $t_0$, of formation.

Note that the region, $R_0$ to R, in which space density is frozen, grows with time. At $R_0$ space density is high – small a(r,t) – and is reduced towards R – higher a(r,t).
At distances r > R, where $\epsilon_{CMB} > \epsilon_g$, space expands.

6. **For galaxies formed ~ 12 BY ago our calculations show that $g_0$ ~ 1.2·10⁻⁸ cm sec⁻².**
This result is based on the value for $\epsilon_{CMB}(R_0, t_0)$, as calculated from its measured value today.
Since $\epsilon_{CMB}$ is distributed homogeneously, we can use equation (2) to obtain the value for $\epsilon_{CMB}(R_0, t_0)$:

$$\epsilon_{CMB}(R_0, t_0) = \epsilon_{CMB}(Now) \cdot \left[\frac{a(Now)}{a(12\,BY\,ago)}\right]^4 = \epsilon_{CMB}(Now) \cdot \frac{1}{a^4}$$

$\epsilon_{CMB}(Now) = 4.17 \cdot 10^{-13}$ erg cm⁻³
Taking 0.26 as the value of a at the time ~12 BY ago (see Section 5 and Perlmutter (2003) p.57) gives: $\epsilon_{CMB}(R_0, t_0) \sim 0.9 \cdot 10^{-10}$ erg cm⁻³
Hence:

$$g_0 = \sqrt{8\pi G \, \epsilon_{CMB}(R_0, t_0)} \sim 1.2 \cdot 10^{-8} \text{ cm sec}^{-2}$$

which, although having the same **value** as the **Millgrom** MOND central acceleration, is **not** a universal constant (M. Millgrom, 2008).

For galaxies formed ~12 BY ago, the constant in the **Tully-Fisher** relation is:
$(g_0 G) \sim 8.7 \cdot 10^{-16}$ cm⁴ sec⁻⁴ gr⁻¹

Note that the data taken from the Perlmutter curves are model-dependent, and hence should be taken just for estimation.

For a given M the distance $R_0$ is:

$$R_0 = \sqrt{\frac{GM}{g_0}}$$

For the Milky Way galaxy, we assume a formation time ~12 BY ago and take M ~ $1.3 \cdot 10^{10}$ $M_\odot$ as the relevant mass, which is the mass of the galaxy's bulge. Our calculation gives $R_0$ ~ 3 KPC, in accordance with observations, (O. Gerhard, 2002). From this distance onwards, an RC is no longer Keplerian, the rotational velocity increases, reaches a plateau and then decreases, as is indicated by observations of dispersion velocities (G. Battagalia, 2005).
Observations show that the MOND central acceleration "universal constant" can take a wide range of values, as our expression for $g_0$ predicts, see K. G. Begeman et al (1991) and D. Scott et al (2001).

To summarize, the gravitational field, $E_g = g$, around a galaxy of mass M, for the three regions of an RC, is:

(12) $r \leq R_0$ $\quad\quad g = \dfrac{GM}{r^2}$ $\quad\quad$ Keplerian RC

(13) $R_0 < r \leq R$ $\quad\quad g = \dfrac{\sqrt{g_0 GM}}{r}$ $\quad\quad$ Flattened RC

In reality, g in the first and second regions depends on the mass distribution and the history of formation.

(14) $r > R$ $\quad\quad g = \dfrac{GM}{r^2}\left[\dfrac{a_z}{a_0}\right]^2$ $\quad\quad$ Keplerian RC

$a_z$ and $a_0$ are defined in Section 3.

7. **The gravitational potential is also modified by space expansion.**
By integrating equation (9) for g, we get the potential difference.

$$\varphi(r) - \varphi(R_0) = \int_{R_0}^{r} g\, dr = \sqrt{g_0(GM)} \cdot \int_{R_0}^{r} r^{-1} dr = \sqrt{g_0(GM)} \cdot \ln\dfrac{r}{R_0}$$

This potential difference is only valid in the region between $R_0$ and R. Since:

$\varphi(R_0) = -\dfrac{GM}{R_0}$ $\quad$ (in reality, $\varphi(R_0)$ depends on the mass distribution) we get:

(15) $\varphi(r) = \sqrt{g_0(GM)} \cdot \ln\dfrac{r}{R_0} - \dfrac{GM}{R_0}$

Fig. (1) illustrates a typical potential in the three regions above as a function of the distance from the center of the galaxy and for different times.

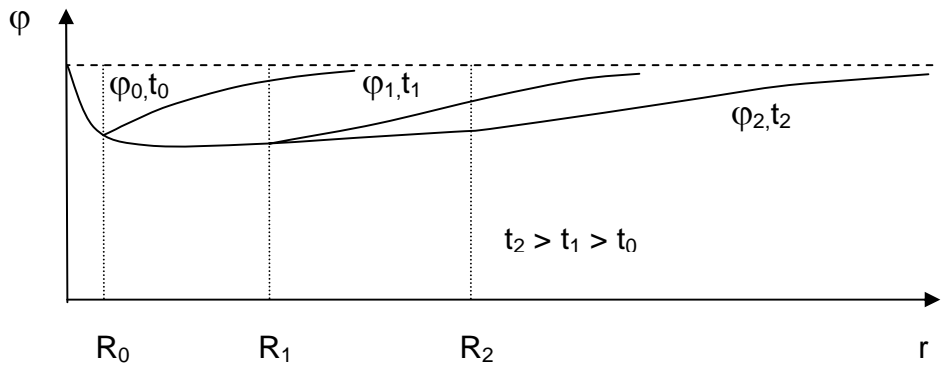

**Fig. (1)  The Gravitational Potential of a Galaxy in an Expanding Universe**

The curves for $\varphi$ in Fig. (1) show that, with time, the zone of flat RC grows. This means that with time "DM halos" should grow. Observations confirm this result.
R. Massey et al (2007) p. 287 observed the evolution of the total DM in the universe.

8. **The gravitational potential in an expanding universe explains the enhanced gravitation lensing.**
   S. M. Carroll (2004) in Sec. 7.3 on Photo Trajectories and Sec. 8.6 on Gravitational Lensing shows that a point mass, M, that serves as a lens deflects a light beam with an impact parameter, b, at the following deflection angle:
   $$\alpha = \frac{4GM}{c^2 b} = \frac{4}{c^2}\varphi$$
   where $\varphi$ is the gravitational potential at a distance b from M.
   However, the potential in the zone of flat RCs around M, expressed by equation (15), yields, for large impact parameters, a much larger deflection of light beams.

9. **"DM Halos" are zones of condensed space with no inertial mass, and hence no gravitational mass.**
   DM halos can be detached from fast moving galaxies like the "bullet cluster" 1E0657-56, (D. Clow et al 2004). We are thus lead to the conclusion that the two ways by which mass deforms space differ from each other as follows:

   - **Elastic deformation by the presence of mass alone.**
     GR shows that space is deformed by gravity, i.e., in the vicinity of masses, space cells are contracted. This contraction is elastic - remove the mass and space resumes its original geometry.

   - **Non-elastic deformation due to space expansion around a mass.**
     In addition to the above elastic deformation, space is also deformed by the inhomogeneous space expansion around the mass, caused by the mass. Such deformation is observed as a DM halo, as shown above. However, in contrast to elastic deformation, the halo does not follow a moving mass and retains its geometry.

We believe that a DM halo, without the presence of a mass, is subject to Hubble expansion although it retains its relative density.

For galaxies, the elastic deformation is orders of magnitude smaller than the non-elastic.

These issues will be discussed in Part III.

Other phenomena attributed to DM halos are explained by our model:

- The larger the initial mass of a galaxy, the larger is $R_0$, and hence we observe Keplerian behavior for $r < R_0$. However, for lower masses, $R_0$ is small and thus space expansion starts close to the center of the galaxy. As a result, we observe an immediate increase in rotational velocity.

- The shapes of DM halos follow the distribution of luminous matter. This is explained by the nature of $R_0$ and R.

## Summary

We consider the effect of the Cosmological Microwave Background on space dilation and its interplay with the gravitational energy density. This explains the inhomogeneous space expansion in, and around, galaxies.

We extend Newton's field equation to account for space deformation caused by this inhomogeneous expansion.

The above leads to a theoretical derivation of the gravitational central acceleration in, and around, galaxies and to a determination of $g_0$, its value where flattening of Rotation Curves (RC) replaces Keplerian behavior.

Our theoretical results fit observations and thus explain the flattening of RCs. This dispels the mystery of Dark Matter.

## Acknowledgements

We would like to thank Roger M. Kaye for his linguistic contribution and technical assistance.

## References

Barak S.  A GeometroDynamic Model (GDM) of Reality (to be published in 2009)
Baugh C. M. et al The Astrophysical Journal 498:504-521 May 10 1998
Battagalia G. et al   arxiv:astro-ph/0506102 v2 10 July 2005
Begeman K. G. et al    Mon. Not. R. astr. Soc 249, 523-537  1991
Carroll S. M.  Spacetime & Geometry  Addison Wesley 2004
Clow D. et al   APJ 604 p. 596 2004
Gerhard O.  arxiv:astro-ph/0203110v1  7 March 2002
Granitt B. R. et al    arxiv:astro-ph/0805.3695V2  10 July 2008
Massey R. et al  Nature Letters Vol 445 18 Jan 2007
Millgrom M.  arxiv:astro-ph/0801.3133v2  3 Mar 2008
Misner C. W. et al   Gravitation 1970  Freeman and Company
Perlmutter S.  Physics Today  April 2003
Scott D. et al   arxiv:astro-ph/0104435v1 26 April 2001
Sakharov A. D.  Soviet Physics – Docklady Vol 12 No. 11 P. 1040 1968